\documentclass[aps,prl,reprint]{revtex4-1} 
\usepackage{graphicx}
\usepackage{amssymb}
\usepackage{epstopdf}
\usepackage{amsfonts}
\usepackage{amsmath}
\usepackage{amssymb}
\usepackage{mathrsfs} 
\usepackage{color}

\newcommand{\pd}[2]{\frac{\partial #1}{\partial #2}}

\newcommand{\ub}[1]{^{({#1})}}

\begin{document}
\title{Extraordinary transmission through a narrow slit}
\author{Jacob R. Holley} \author{Ory Schnitzer}
\affiliation{Department of Mathematics, Imperial College London, SW7 2AZ, UK}
\begin{abstract}
We revisit the problem of extraordinary transmission of acoustic (electromagnetic) waves through a slit in a rigid (perfectly conducting) wall. We use matched asymptotic expansions to study the pertinent limit where the slit width is small compared to the wall thickness, the latter being commensurate with the wavelength. Our analysis focuses on near-resonance frequencies, furnishing elementary formulae for the field enhancement, transmission efficiency, and deviations of the resonances from the Fabry--P\'erot frequencies of the slit. We find that the apertures' near fields play a dominant role, in contrast with the prevalent approximate theory of Takakura [\textit{Physical Review Letters}, \textbf{86} 5601 (2001)]. Our theory agrees remarkably well with numerical solutions and electromagnetic experiments [Suckling \textit{et al.}, \textit{Physical Review Letters}, \textbf{92} 147401 (2004)], thus providing a paradigm for analyzing a wide range of wave propagation problems involving small holes and slits. 
\end{abstract}
\maketitle

\textit{Introduction}.---The term extraordinary transmission originated in electromagnetism \cite{Ebbesen:98} to describe enhanced transmission of wave energy through small apertures via excitation of surface plasmons, spoof plasmons and localized resonances, with analogous effects identified in acoustics \cite{Christensen:08} and water waves \cite{Evans:18}. An elementary example, that has been widely studied experimentally and theoretically, is transmission of TM-polarized electromagnetic waves through a single slit in a metallic wall whose thickness is comparable to the wavelength and large compared to the slit width \cite{Takakura:01,Yang:02,Garcia:03,Suckling:04,Lin:17}. When the metal is perfectly conducting then the problem is analogous to (lossless) acoustic transmission through a slit in a rigid wall \cite{Christensen:08,Ward:15}. In these idealized settings enhanced transmission is attributed to excitation of standing waves in the slit, or Fabry--P\'erot resonances, which leak energy via diffraction at the slit ends; with decreasing slit width, the on-resonance transmission efficiency and slit-field magnitude are enhanced while the transmission peaks approach the standing-wave frequencies.

Takakura \cite{Takakura:01} was the first to put forward an approximate theory of ideal transmission through a narrow slit, based on an \textit{ad hoc} truncation of an exact mode-matching scheme. Takakura's key result, a simple closed-form approximation for the deviations of the transmission peaks from the Fabry--P\'erot resonances, was found to agree poorly with electromagnetic experiments \cite{Suckling:04}; the discrepancy could not be solely attributed to material losses or a skin effect as it persisted even for slits wide enough for the metal to be considered perfectly conducting (still exceedingly narrow relative to the wall thickness) \footnote{Earlier, less accurate, experiments  \cite{Yang:02} seemed to show good agreement with Takakura's approximation.}.  
More recent attempts to derive rigorous approximations for ideal transmission through a single slit \cite{Joly:06b,Lin:17} have not provided explicit formulae or physical insight into the discrepancy. 

In this Letter we systematically develop an approximate theory of extraordinary transmission through a narrow slit. Our theory is based on an asymptotic analysis of the ideal transmission problem (described in the language of acoustics, for convenience) in the pertinent limit where the slit width is small compared to the wall thickness, the latter being comparable to the wavelength. We demonstrate that our results, which include asymptotic formulae for the transmission efficiency, field enhancement and frequency shifts, are in excellent agreement with numerical and experimental data; moreover, we expose the logarithmic error rendering Takakura's formula inaccurate and point out the (very common) flawed physical assumption at the basis of that approximation. 

The notably simple form of our theory follows from the manner in which we treat the disparate length scales and distinguished frequency regimes in the problem. Namely, rather than attempting to find a single approximation for the wave field that is uniformly valid everywhere and for all frequencies, we use the method of matched asymptotic expansions \cite{Hinch:91} to systematically decompose the physical domain into asymptotically overlapping regions (representing the slit, the external regions and transition regions near the apertures) and separately consider off- and near-resonance frequencies; our analysis is thereby distinct from  recent asymptotic analyses of related transmission problems \cite{Evans:17,Schnitzer:17,Evans:18,Porter:18}. 

\begin{figure}[t!]
	\begin{center}
		\includegraphics[scale=0.5]{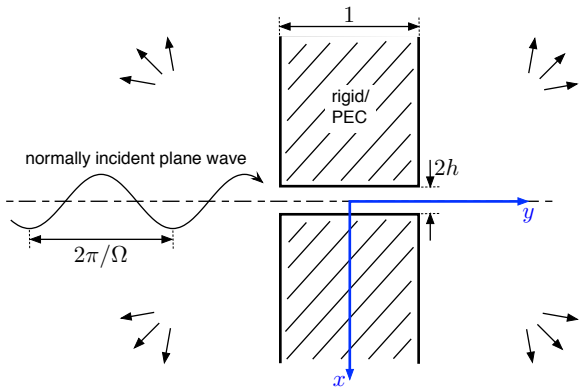}
		\caption{Schematic of the dimensionless transmission problem. }
		\label{fig:schematic}
	\end{center}
\end{figure}

\textit{Formulation}.---Consider acoustic transmission of a normally incident plane wave (angular frequency $\omega$, wave speed $c$) through a slit of width $2lh$ in a rigid wall of thickness $l$. We suppress the time dependence $e^{-i\omega t}$ in the usual way and adopt a dimensionless formulation where lengths are normalised by $l$ and $\varphi$ denotes the velocity potential normalised by the amplitude of the incident wave. 
Figure \ref{fig:schematic} shows a dimensionless schematic of the problem and defines the Cartesian coordinates $(x,y)$. 

In the fluid domain $\varphi$ satisfies the Helmholtz equation
\begin{equation}\label{helm}
\nabla^2\varphi+\Omega^2\varphi=0,
\end{equation}
where we define the dimensionless frequency
\begin{equation}\label{freq}
\Omega=\frac{\omega l}{c}.
\end{equation}
On the walls, impermeability implies that the normal derivative vanishes,
\begin{equation}\label{bc}
\pd{\varphi}{n}=0.
\end{equation}
In addition, the scattered field $\varphi-\varphi\ub{i}$, where $\varphi\ub{i}=e^{i\Omega y}$ is the incident wave, must propagate outwards at large distances from the slit. (In the electromagnetic analogy mentioned in the introduction, the same dimensionless problem holds with $\varphi$ replaced by the scaled out-of-plane magnetic-field component.)

In what follows we shall be interested in the resonant peaks of the transmission efficiency $\eta$, defined as the ratio of the acoustic (or electromagnetic) power transmitted through the slit to that transmitted through the same cross section in the absence of a wall:
\begin{equation} \label{ap}
\eta \equiv \frac{1}{2h\Omega } \int_{-h}^{h} \mathrm{Im}\left\{\varphi^* \pd{\varphi}{y}\right\} \, dx,
\end{equation}
where the asterisk denotes complex conjugation and the integrand is evaluated for fixed $y$  such that $|y|<1/2$.

\begin{figure}[t!]
	\begin{center}
		\includegraphics[scale=0.45]{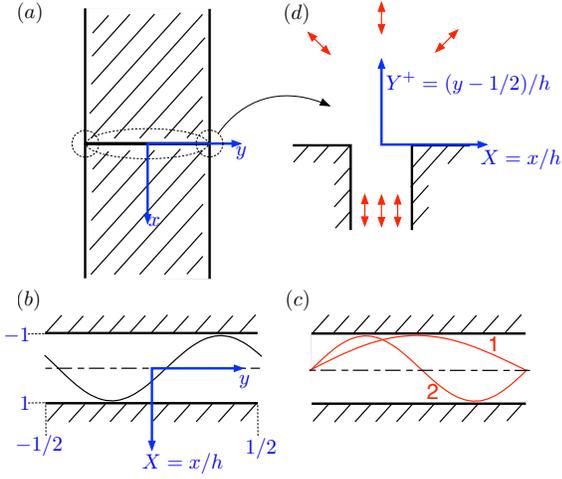}
		\caption{Asymptotic outer regions (a) and inner slit region (b). Near resonances, the slit field is approximately a Fabry--P\'erot mode enhanced by an $\mathcal{O}(h^{-1})$ factor (c) with the transition regions (d) close to the apertures playing a dominant role.}
		\label{fig:regions}
	\end{center}
\end{figure}

\textit{Analysis off-resonance}.---Let us first, naively, consider the narrow-slit limit $h\ll1$ holding $\Omega$ fixed. The slit openings reduce to points, as depicted in figure \ref{fig:regions}(a). Assuming that the slit potential is comparable in magnitude to the incident wave, as we shall later verify, diffraction from the slit is negligible in the domain left of the wall; the potential there, say $\varphi^-$, is accordingly 
\begin{equation}\label{left naive}
\varphi^-\approx e^{i\Omega y}+R e^{-i\Omega y},
\end{equation}
where the reflection coefficient $R=e^{-i\Omega}$ follows from \eqref{bc} and $\approx$ denotes a relative asymptotic error of algebraic order, i.e., scaling with some positive power of $h$. Right of the wall, there is no incident nor reflected wave, only weak diffraction from the slit; the potential there, $\varphi^+$, is accordingly algebraically small. 

The above approximations are outer expansions, namely with position held fixed. We next derive an inner approximation for the wave field inside the slit by considering the limit $h\to0$ with the stretched transverse coordinate $X=x/h$ fixed. Defining the slit field $\varphi(x,y)=\Phi(X,y)$, for $|y|<1/2$, it is straightforward to show from a regular perturbation of \eqref{helm} and \eqref{bc} that $\Phi\approx U(y)$, where $U(y)$ satisfies
\begin{equation}\label{1d wave}
\frac{d^2U}{dy^2}+\Omega^2 U =0
\end{equation}
with solution (see figure \ref{fig:regions}(b))
\begin{equation}\label{U naive}
U(y)=A\cos(\Omega y) + B \sin(\Omega y).
\end{equation}
To determine the constants $A$ and $B$ we asymptotically match the slit field to the outer domains; noting that \eqref{left naive} tends to a limiting value as $y\to-1/2$ and that $\varphi^+$ is asymptotically small, we find
\begin{equation}\label{naive AB}
A=e^{-i\Omega/2}\sec\frac{\Omega}{2}, \quad B  = -e^{-i\Omega/2}\csc\frac{\Omega}{2}.
\end{equation}
Based on this solution the dimensionless flux density in the $y$ direction, $\partial\varphi/\partial y$, is $O(1)$ in the slit; at the right end of the slit the corresponding net flux is $\approx 2hU'(1/2)$. This flux is approximately conserved on the small scale of the aperture, implying that the outer potential $\varphi^+$ is given by a transmitted cylindrical wave \footnote{Higher-order fundamental solutions must be discarded as their algebraic singularity contradicts the order of magnitude of the slit potential.}
\begin{equation}\label{naive outer right}
\varphi^+ \approx hQ\mathcal{H}_0(\Omega r^+),
\end{equation}
where $\mathcal{H}_0$ is the zeroth-order Hankel function, $r^{\pm}=\sqrt{x^2+(y\mp1/2)^2}$ and 
\begin{equation}\label{naive Q}
Q=2ie^{-i\Omega/2}\Omega\csc\Omega.
\end{equation}
Here and later we use the asymptotic relation
\begin{equation}
\mathcal{H}_0(t) \approx \frac{2i}{\pi}\left(\ln\frac{t}{2}+\gamma_E\right)+1  \quad \text{as} \quad t\to0,
\end{equation}
wherein the algebraic error is in $t$ and $\gamma_E\approx0.5772\ldots$ is the Euler--Mascheroni constant \cite{Abramowitz:book}.

Our fixed-frequency approximation gives no pronounced enhancement, \textit{viz.} $\eta=O(1)$, except near the Fabry--P\'erot frequencies
\begin{equation}\label{FP}
\bar{\Omega}=m\pi, \quad m=1,2,\ldots,
\end{equation}
where \eqref{naive AB} and \eqref{naive Q} are singular; as $\Omega\to\bar\Omega$ the slit field $\mathcal{U}(y)$ diverges in amplitude and approaches a standing-wave solution vanishing at $y=\pm1/2$. The singular nature of the fixed-frequency limit is evident on physical grounds: as diffraction from the slit ends is neglected (and material losses have been discarded from the outset), the limiting standing waves constitute  resonant modes. For $\Omega-\bar\Omega=O(h)$, however, our present approximation suggests that the amplitude of the slit field is $O(1/h)$ and the diffracted waves are $O(1)$, comparable to the incident and reflected waves. Our asymptotic analysis therefore needs to be modified in this regime. 

\textit{Analysis near resonance}.---In light of the above, we now define
\begin{equation}
\frac{\Omega-\bar\Omega}{\bar\Omega}\equiv 2h \Omega'
\end{equation}
and in what follows consider the near-resonance limit $h\to0$ with $\Omega'$ (and $m$) fixed. 
In this modified limit we anticipate the diffracted waves in the outer regions to be $O(1)$, i.e.,
\begin{equation}\label{left enhanced}
\varphi^-\approx e^{i\bar{\Omega} y}+e^{-i\bar{\Omega}(1+y)}+Q^-\mathcal{H}_0\left(\bar{\Omega} r^-\right)
\end{equation}
and
\begin{equation}\label{right enhanced}
\varphi^+\approx Q^+\mathcal{H}_0\left(\bar{\Omega} r^+\right),
\end{equation}
where the coefficients $Q^\pm$ remain to be determined. The slit potential is accordingly amplified, 
\begin{equation}\label{enhanced slit}
\Phi(X,y) \approx h^{-1}\mathcal{U}(y),
\end{equation}
where $\mathcal{U}(y)$ satisfies equation \eqref{1d wave} with $\bar{\Omega}$ replacing $\Omega$, whereas the relatively negligible magnitude of the outer potentials \eqref{left enhanced} and \eqref{right enhanced} implies the boundary conditions $\mathcal{U}(\pm 1/2)=0$. The resulting homogeneous problem for $\mathcal{U}(y)$, by construction, has as non-trivial solutions 
\begin{gather}\label{homo solutions}
	\mathcal{U}(y) = \mathcal{A} \times \left\{\begin{array}{c}\cos(\bar{\Omega} y ) \\ \sin(\bar{\Omega} y )\end{array}\right\},
\end{gather}
where $\mathcal{A}$ is a prefactor and henceforth the upper element of an array corresponds to odd $m$ (even standing wave) and the lower to even $m$ (odd standing wave). 

The analysis of the slit region can be readily extended one further algebraic order by writing \begin{equation}\label{slit higher order}
\Phi(X,y) \approx h^{-1}\mathcal{U}(y)+\mathcal{V}(y), 
\end{equation}
where $\mathcal{V}(y)$ satisfies 
\begin{equation}\label{correction}
\frac{d^2\mathcal{V}}{dy^2}+\bar{\Omega}^2 \mathcal{V} =-4\Omega'\bar{\Omega}^2\mathcal{U}.
\end{equation}
(It is straightforward to verify that the slit potential deviates from a unidimensional profile only at higher algebraic order.) Multiplying equation \eqref{correction} by $\mathcal{U}^*$ and subtracting the conjugate of equation \eqref{1d wave} (with $\bar{\Omega}$ replacing $\Omega$) multiplied by $\mathcal{V}$, followed by integration along the slit using $\mathcal{U}(\pm 1/2)=0$, gives
\begin{equation}\label{pre solvability}
\left[\mathcal{V}\frac{d\mathcal{U}^*}{dy}\right]_{-1/2}^{1/2}=4\Omega'\bar{\Omega}^2\int_{-1/2}^{1/2}|\mathcal{U}|^2\,dy.
\end{equation}
Substituting \eqref{homo solutions} we find 
\begin{equation}\label{sc}
\mathcal{A}= \frac{i^m}{2\Omega'\bar\Omega}\times \left\{\begin{array}{c}i(\mathcal{V})_{1/2}+i(\mathcal{V})_{-1/2} \\ (\mathcal{V})_{1/2}-(\mathcal{V})_{-1/2} \end{array}\right\}.
\end{equation}

\begin{figure}[t!]
	\begin{center}
		\includegraphics[scale=0.5]{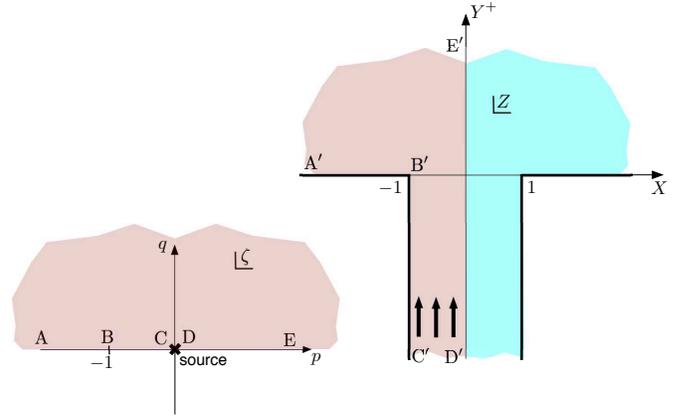}
		\caption{Conformal mapping of the left half ($X < 0$) of the transition region into the upper half of an auxiliary $\zeta$ plane.}
		\label{fig:map}
	\end{center}
\end{figure}
As in the off-resonance analysis, the leading fluxes at the slit ends can be matched with the corresponding diffraction terms in the outer regions; we thereby find
\begin{equation}\label{flux relation res}
Q^{\pm} = i^m\bar{\Omega}\mathcal{A}\times \left\{\begin{array}{c} 1 \\ \mp i \end{array}\right\}.
\end{equation}
Unlike in the off-resonance analysis, however, the slit and outer potentials cannot be directly matched, since the leading-order outer potentials are singular as $r^\pm\to0$. It is therefore necessary to consider transition regions near the apertures defined by intermediate limits with $X=x/h$ and $Y^\pm=\pm(y\mp1/2)/h$ fixed (we also define $R^{\pm}=r^{\pm}/h$); under these scalings, the slit appears infinite and the boundaries are $Y^\pm=0$ for $|X|>1$ and $X=\pm1$ for $Y^{\pm}<0$ (see figure \ref{fig:regions}(d)). It is evident from the form of the slit and outer expansions that the potential is $O(1)$ in the transitions regions; thus let $\varphi(x,y) \approx T^{\pm}(X,Y^\pm)$, where $T^{\pm}$ satisfies Laplace's equation
\begin{equation}\label{T eq}
\pd{^2T^{\pm}}{X^2}+\pd{^2T^{\pm}}{Y^{\pm2}}=0,
\end{equation}
the Neumann boundary condition 
\begin{equation}\label{T Normal}
\pd{T^\pm}{N}=0,
\end{equation}
where $N$ denotes $Y^{\pm}$ for the external walls and $X$ for the inner slit walls, as well as asymptotic matching with the outer and slit regions. In the off-resonance limit the $O(1)$ outer and slit potentials were regular hence $T^{\pm}$ were constants; in the present near-resonance limit, however, matching with the outer expansions 
gives
\begin{multline}\label{T far R}
T^{\pm} \sim iQ^{\pm}\frac{2}{\pi}\ln R^{\pm}+Q^{\pm}\left[1+\frac{2i}{\pi}\left(\ln\frac{\bar{\Omega} h}{2}+\gamma_E\right)\right]\\
+(1\mp 1)e^{-i\bar{\Omega}/2} + o(1) \quad \text{as} \quad R^{\pm}\to\infty,
\end{multline}
while matching with the slit expansion gives
\begin{equation}\label{T far Y}
T^{\pm}\sim iQ^{\pm}Y^{\pm}+(\mathcal{V})_{\pm 1/2} +o(1) \quad \text{as} \quad Y^{\pm} \to-\infty, 
\end{equation}
where the leading term in \eqref{T far Y} is written in terms of $Q^{\pm}$ using \eqref{flux relation res}. 
\begin{figure}[t!]
	\begin{center}
		\includegraphics[scale=0.32]{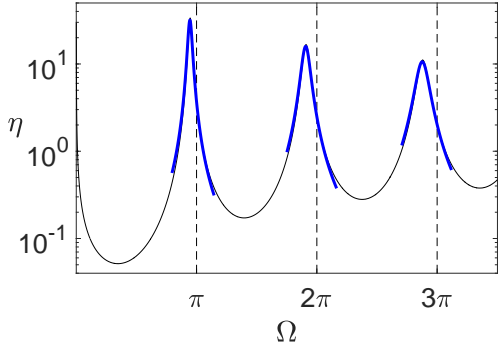}
		\caption{Numerical (black line) and near-resonance asymptotic (thick blue lines) transmission efficiency, for $h=0.01$, plotted as a function of the normalized frequency.}		
		\label{fig:trans}
	\end{center}
\end{figure}

The Laplace problem for $T^{+}$ (similarly $T^-$) is  uniquely solvable, up to an arbitrary additive constant, with just the leading terms in the far-field conditions \eqref{T far R} and \eqref{T far Y} prescribed; that unique solution therefore determines the differences between the $O(1)$ constant terms in those conditions. To treat both Laplace problems simultaneously we note that $T^{\pm}/iQ^{\pm}$ solves a joint canonical  problem where the difference 
\begin{multline}\label{beta equiv}
\beta \equiv \lim_{Y^{\pm}\to-\infty}\left(\frac{T^{\pm}}{iQ^{\pm}}-Y^{\pm}\right)\\-\lim_{R^{\pm}\to\infty}\left(\frac{T^{\pm}}{iQ^{\pm}}  -\frac{2}{\pi}\ln R^{\pm}\right)
\end{multline}
is a pure number, which depends only on the geometry of the aperture. In the present geometry we obtain $\beta$  by defining the complex variable $Z=X+iY^{+}$ and considering  the conformal mapping \cite{Churchill:Book}
\begin{equation}\label{mapping}
Z = \frac{2i}{\pi}(1+\zeta)^{1/2}+\frac{i}{\pi}\log\frac{(1+\zeta)^{1/2}-1}{(1+\zeta)^{1/2}+1}
\end{equation}
from the upper half-plane of an auxiliary complex variable $\zeta=p+iq$ to the left half $(X<0)$ of the $XY^+$ domain (see figure \ref{fig:map}). The form of the solution is evident in terms of the auxiliary variable:
\begin{equation}\label{mapped solution}
\frac{T^+}{iQ^+}=\frac{1}{\pi}\ln|\zeta| + \text{const.}
\end{equation}
Using \eqref{mapped solution} and the limits of \eqref{mapping} as $\zeta\to0$ and $\zeta\to\infty$, definition \eqref{beta equiv} yields 
\begin{equation}
\beta = \frac{2}{\pi}\left(\ln\frac{4}{\pi}-1\right).
\end{equation}

With $\beta$ determined, the complex amplitude of the slit wave is found from \eqref{sc}, \eqref{flux relation res} and \eqref{T  far R}--\eqref{beta equiv} as 
\begin{equation}\label{A}
\mathcal{A}=\frac{1/\bar\Omega}{\Omega'-\frac{2}{\pi}\left(\ln\frac{2\bar{\Omega}h}{\pi}+\gamma_E-1\right)+i}\times \left\{\begin{array}{c}i \\ -1 \end{array}\right\}.
\end{equation}
From this main result we find the transmission efficiency, 
\begin{equation}\label{eff}
\eta \approx  h^{-1}\bar\Omega|\mathcal{A}|^2,
\end{equation}
and field enhancement in the slit, $|\mathcal{A}|/h$, in terms of the amplitude 
\begin{equation}\label{A amplitude}
|\mathcal{A}|^2= \frac{(\pi m)^{-2}}{1+\left[\Omega'-\frac{2}{\pi}\left(\ln(2mh)+\gamma_E-1\right)\right]^2},
\end{equation}
a Lorenzian with a peak magnitude decreasing with $m$; the frequency deviations $\Delta f$, of the resonant peaks from the Fabry--P\'erot frequencies $f_m=mc/(2l)$, are 
\begin{equation}\label{shifts}
\frac{\Delta f}{f_m}\approx\frac{4h}{\pi}\left(\ln h + \ln(2m) +\gamma_E-1\right).
\end{equation}

\begin{figure}[t!]
	\begin{center}
		\includegraphics[scale=0.24]{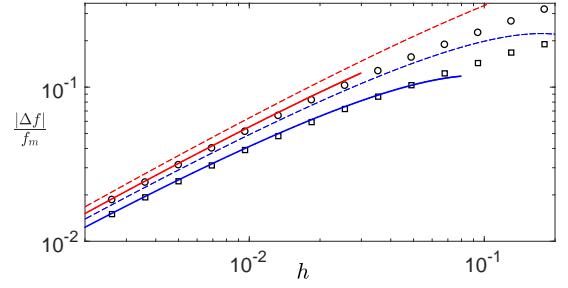}
		\caption{Relative frequency shifts from Fabry-P\'erot values (numerics: symbols; asymptotics: solid lines) as a function of $h$ for modes $m=1$ (circles) and $m=3$ (squares). The dashed lines depict Takakura's approximation \eqref{shifts taka}.}
		\label{fig:shifts}
	\end{center}
\end{figure}
\begin{figure}[t!]
	\begin{center}
		\includegraphics[scale=0.25]{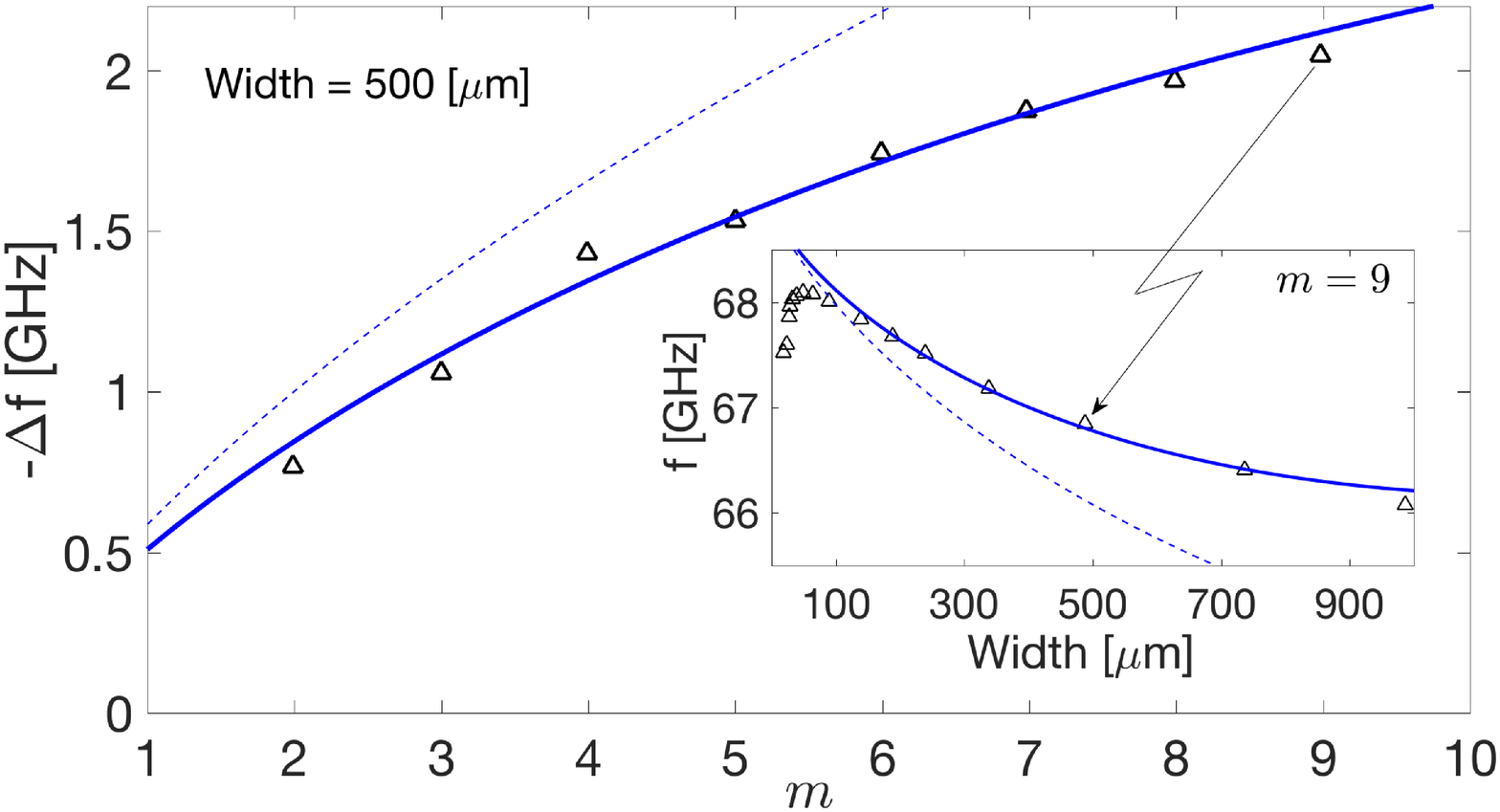}
		\caption{Comparison with experiments of Ref.~\cite{Suckling:04} of electromagnetic transmission through a slit in a $19.58mm$ thick Aluminium plate (triangles --- using plot digitizer). Main plot shows shifts of resonance frequencies from the second to ninth Fabry-P\'erot mode for a $500 \mu m$ wide slit. Inset shows the resonance frequencies for mode $m=9$ for varying slit width. In both we depict our asymptotic prediction by a solid line and Takakura's approximation \eqref{shifts taka} by a dashed line.}	
\label{fig:exp}
	\end{center}
\end{figure}

\textit{Discussion and concluding remarks}.---It is illuminating to compare our asymptotic theory with numerical calculations, experimental data and existing approximations. To begin with, we have validated our theory against a mode-matching scheme solving the ideal transmission problem exactly \footnote{Please contact authors for the supplementary information.}; in particular, figures \ref{fig:trans} and \ref{fig:shifts} present excellent agreement between our on-resonance asymptotic predictions for $\eta$ and $\Delta f$ [cf.~\eqref{eff}--\eqref{shifts}] and the corresponding numerical values.
In figure \ref{fig:shifts} we have added a dashed line depicting Takakura's prediction for the frequency shifts \cite{Takakura:01}:
\begin{equation}\label{shifts taka}
\left(\frac{\Delta f}{f_m}\right)_{\text{Ref.~\cite{Takakura:01}}} = \frac{4h}{\pi}\left(\ln h + \ln (\pi m) -\frac{3}{2}\right)
\end{equation}
(ignoring terms of quadratic order in $h$). Takakura's approximation is seen to be in poor agreement with the numerical data. Note that \eqref{shifts} and \eqref{shifts taka} disagree at $O(h)$ which is practically comparable to the leading $O(h\ln h)$ term \footnote{The leading $O(h\ln h)$ term was also found by Lin \textit{et al.} \cite{Lin:17}. Their expression for the $O(h)$ term, however, includes a constant which was not determined.}; our analysis suggests that the logarithmic error in Takakura's approximation is a result of his disregard of the details of the wave field close to the apertures, which are seen to be important specifically near resonance. Finally, in figure \ref{fig:exp} we replicate the comparison carried out by Suckling \textit{et al.} \cite{Suckling:04} between their electromagnetic measurements and Takakura's approximation \eqref{shifts taka}, with our asymptotic prediction \eqref{shifts} overlaid. Our approximation is seen to be in excellent agreement with the data in an intermediate range of slit widths: narrow enough so that $h \ll 1$ and wide enough such that the metal can be modelled as a perfect conductor to a good approximation. 

In conclusion, our asymptotic analysis provides a simple, accurate and physically representative theory of ideal extraordinary transmission through a single narrow slit, which in particular improves agreement with experiments and highlights the importance of aperture effects close to resonance. As the slit width is reduced, holding the wall thickness and frequency fixed, intrinsic (material) losses ultimately become dominant over radiation damping. 
This is evident in the electromagnetic experimental data of Ref.~\cite{Suckling:04} shown in the inset to figure \ref{fig:exp} and more pronouncedly in recent acoustic experiments \cite{Ward:15, Ward:16}. It is therefore desirable to extend the present theory to incorporate losses, which are of an inherently different nature in the electromagnetic and acoustic cases \cite{Suckling:04,Ward:15,Moleron:16}. Our asymptotic approach may offer simplification in both of these non-ideal regimes, which are fundamental to the practical design of acoustic and photonic metamaterials.
OS is grateful for support from EPSRC through grant EP/R041458/1. 

\bibliographystyle{apsrev4-1}
\bibliography{refs}

\end{document}